\documentclass[11pt,a4paper,twoside]{article}
\usepackage{amssymb,amsfonts,amsmath,amsthm}
\usepackage[T1]{fontenc}
 \usepackage[english]{babel}
\usepackage[left=2.7cm,right=2.7cm,bottom=2.67cm,top=2.67cm]{geometry}
\usepackage{dsfont}
\usepackage{mathrsfs}
\usepackage{natbib}
\usepackage{enumerate}
\usepackage{color}
\RequirePackage[OT1]{fontenc}
\usepackage{graphicx}
\usepackage{multirow}
\usepackage{multicol}
\usepackage{subfigure}
\usepackage{placeins}
\usepackage{rotating}
\usepackage[flushleft]{threeparttable}
\usepackage{booktabs}
\usepackage{xcolor}
\usepackage[normalem]{ulem}
\usepackage[implicit=false]{hyperref}

\graphicspath{{Figures/}}

\theoremstyle{plain}

\theoremstyle{definition}
\newtheorem{remark}{Remark}[section]

\newcommand{\F}{\mathscr{F}}
\newcommand{\R}{\mathds{R}}

\newcommand{\Z}{\mathds{Z}}
\newcommand{\eps}{\varepsilon}
\newcommand{\bs}[1]{\boldsymbol{#1}}

\long\def\sfootnote[#1]#2{\begingroup%
\def\thefootnote{\fnsymbol{footnote}}\footnote[#1]{#2}\endgroup}

\def\bfootnote{\xdef\@thefnmark{}\@footnotetext}

\begin{document}
\pagestyle{myheadings} 
\markboth{Bayes Estimation of GLARMA Models With Applications}{G. Pumi, A.J. Camara}

\thispagestyle{empty}
{\centering
\Large{\bf Bayes Estimation of GLARMA Models With Applications}\vspace{.5cm}\\

\normalsize{ {\bf Guilherme Pumi${}^{\mathrm{a,}}$\sfootnote[1]{Corresponding author. This Version: \today},\let\thefootnote\relax\footnote{\hskip-.3cm$\phantom{s}^\mathrm{a}$Mathematics and Statistics Institute and Programa de P\'os-Gradua\c c\~ao em Estat\'istica - Universidade Federal do Rio Grande do Sul.
} Ana Julia Alves Camara${}^\mathrm{b}$ \let\thefootnote\relax\footnote{\hskip-.3cm$\phantom{s}^\mathrm{b}$Universidade Federal do Esp\'irito Santo.}
 \\
\let\thefootnote\relax\footnote{E-mails: guilherme.pumi@ufrgs.br (G. Pumi), anajulia.camara@gmail.com (A.J.A. Camara) \\ 
\indent \quad ORCID's: 0000-0002-6256-3170 (Pumi); 0000-0002-9382-6842 (Camara)}\\
\vskip.3cm
}}}

\begin{abstract}
This work presents a Bayesian approach for parameter estimation in the class of Generalized Linear Autoregressive Moving Average (GLARMA) models, extending the methodology beyond the common exponential family setting. The proposed framework accommodates positive, double-bounded, and count time series through a unified MCMC-based estimation procedure implemented in \texttt{nimble}. We discuss prior specifications for the model parameters and conduct an extensive Monte Carlo simulation study to evaluate the finite-sample performance of the approach under three distinct data-generating mechanisms: Negative Binomial-GLARMA for count data, Beta-GLARMA for double-bounded outcomes, and Gamma-GLARMA for positive continuous time series. The simulation study assesses point and interval estimation, prior sensitivity, and the behaviour of the estimators under varying levels of temporal persistence, including challenging scenarios near the boundaries of the stationarity region. The practical utility of the proposed framework is illustrated through two empirical applications: analysing monthly net electricity generation by nuclear plants in the United States using a Gamma-GLARMA model with harmonic seasonal components; and modelling monthly hospital admissions due to chronic obstructive pulmonary disease in Belo Horizonte, Brazil, using a Negative Binomial-GLARMA model with principal components derived from air pollution covariates. The results demonstrate that the proposed Bayesian framework provides reliable and stable estimation across all settings, offering a flexible and practical tool for analysing non-Gaussian time series in a wide range of applications.\\[.2cm]
\end{abstract}

\section{Introduction}
Environmental and energy time series frequently exhibit features that challenge conventional Gaussian modelling approaches. Data such as air pollution concentrations, streamflow levels, and energy generation are naturally constrained to positive values; relative humidity, vegetation cover indices, and mortality or morbidity rates are double-bounded in the unit interval; while counts of adverse health events, species abundance, or extreme weather occurrences are inherently discrete. Although Gaussian linear models and classical ARMA processes are sometimes employed for such data, their use carries tangible consequences. Perhaps the most well-known drawback is that out-of-sample forecasts can fall outside the natural bounds of the data, producing physically implausible predictions. A clear illustration is provided by \cite{grande}, who modelled influenza dynamics in Brazil using Gaussian linear regression, focusing on peak prediction; while useful for that purpose, their approach yielded negative forecasts for valleys (regions of no interest in that study).

Environmental and energy time series frequently exhibit features that challenge conventional Gaussian modelling approaches. Data such as air pollution concentrations, streamflow levels, relative humidity, and energy generation are naturally constrained to positive or bounded intervals, while counts of adverse health events, species abundance, or extreme weather occurrences are inherently discrete. Although Gaussian linear models and classical ARMA processes are sometimes employed for such data, their use carries tangible consequences. Perhaps the most well-known drawback is that out-of-sample forecasts can fall outside the natural bounds of the data, producing physically implausible predictions. A clear illustration is provided by \cite{grande}, who modelled influenza dynamics in Brazil using Gaussian linear regression, focusing on peak prediction; while useful for that purpose, their approach yielded negative forecasts for valleys (regions of no interest in that study).

A related and particularly perilous practice is data transformation. It is well documented that transformations can drastically alter the dependence structure of a time series \citep{Sangs}, and should therefore be avoided whenever possible. Given the pitfalls of both Gaussian approximations and transformations, it is natural to seek models that respect the data's inherent scale. The class of Generalized Linear Autoregressive Moving Average (GLARMA) models \citep{Davis1999,Davis2003}, offers such a solution. By embedding ARMA dynamics into a generalized linear model framework, GLARMA provides a principled and flexible approach that naturally accommodates bounded, double-bounded, and count outcomes, as well as covariates and asymmetry, unlike generic non-Gaussian state-space models, which often sacrifice interpretability.

The environmental sciences present numerous settings where such modelling flexibility is essential. Hydrological variables such as rainfall, streamflow, and reservoir levels are inherently positive and often exhibit pronounced persistence and seasonal patterns \citep{Peitzsch2021}. Air quality indicators, including particulate matter concentrations and ozone levels, are positive-valued and strongly influenced by meteorological conditions and emission sources. Counts of hospital admissions due to respiratory or cardiovascular diseases are discrete and depend on complex interactions between air pollution, weather, and socioeconomic factors \citep{Camara2025}. Similarly, energy systems generate data that are inherently positive, such as electricity production and consumption, which are subject to daily, weekly, and seasonal cycles, as well as long-term trends and policy interventions.

The GLARMA literature has grown steadily in recent years, both in applications \citep{Peitzsch2021, Camara2025, CamaraAMMOD} and methodological extensions \citep{Franco2019, Maia2025, Maia2026}. Parameter estimation has been conducted predominantly under a frequentist paradigm via conditional maximum likelihood. Bayesian approaches, in contrast, have received considerably less attention. To the best of our knowledge, \cite{Franco2019} is the only work to consider a Bayesian treatment of GLARMA models, and that contribution is restricted to count time series. This gap is particularly relevant for environmental applications, where prior information from historical data, physical knowledge, or expert elicitation can be naturally incorporated to improve estimation and prediction, especially in settings with limited sample sizes.

In this paper, we aim to fill this gap by developing a fully Bayesian estimation framework for general GLARMA models, beyond those whose random component belongs to the exponential family. Our approach enables Bayesian inference for positive, double-bounded, and count time series, as well as for any other response type that can be accommodated within the GLARMA structure. This generalization is particularly important because in many environmental applications, distributions outside the exponential family, such as two-parameter Gamma, Beta, and Negative Binomial, 
are better suited for inference, where existing frequentist methods may encounter numerical difficulties or lack a principled treatment of parameter uncertainty.

To evaluate the performance of the proposed framework in finite samples and to showcase different data-generating mechanisms relevant to environmental and energy systems, we conduct a simulation study considering Negative Binomial-GLARMA for count data, Beta-GLARMA for $(0,1)$-valued outcomes, and Gamma-GLARMA for positive continuous time series. We assess point and interval estimation, prior sensitivity, and the behaviour of the estimators under varying levels of persistence, including challenging scenarios near the boundaries of the stationarity region. 

Finally, to illustrate the practical utility of our approach in environmental and energy contexts, we present two empirical applications. The first analyses monthly net electricity generation by nuclear plants in the United States using a Gamma-GLARMA model with harmonic seasonal components, demonstrating how the proposed framework can capture the positive-valued, persistent, and seasonally varying nature of energy production data. The second models monthly hospital admissions due to chronic obstructive pulmonary disease in Belo Horizonte, Brazil, using a Negative Binomial-GLARMA model with principal components derived from air pollution covariates, showcasing the framework's ability to handle discrete responses and high-dimensional covariate information in environmental health studies. These applications highlight the flexibility of our Bayesian framework in handling time series from diverse environmental domains within a unified modelling paradigm.

The remainder of this paper is organized as follows. Section 2 formally introduces the class of GLARMA models, discussing its observation-driven structure, the recursion defining the systematic component, and the conditional likelihood framework that underpins both frequentist and Bayesian estimation. Section 3 presents our proposed Bayesian inference procedure, including prior specifications and the MCMC implementation. Section 4 describes the Monte Carlo simulation study, evaluating the performance of the approach under Negative Binomial, Beta, and Gamma GLARMA specifications. Section 5 presents the empirical applications to U.S. nuclear energy generation data and COPD hospital admissions in Brazil. Section 6 concludes with a discussion of our findings and outlines directions for future research.

\section{GLARMA Models}

Let $\{Y_t\}_{t\in\Z}$ be a time series of interest satisfying, and let $\{\bs X_t\}_{t\in\Z}$ denote a set of $r$-dimensional exogenous time dependent (possibly random) covariates. Let $\F_{t}$ denote the information ($\sigma$-field) available to the observer at time $t$, that is,   $\F_{t}:=\sigma\{\bs X_{t+1}, Y_{t},\bs X_{t},Y_{t-1},\cdots\}$, where, by convention, $\bs X_t$ denotes the observed values at time $t$ for deterministic covariates, and at time $t-1$ for stochastic ones. GLARMA are observation-driven models \citep[see][]{Davis1999}; in this framework, the parameters specifying the stochastic process are deterministically determined by its past observations. As such, these models are defined by specifying the so-called random and systematic components. For the random component, we assign a probability density or mass function to $Y_t|\F_{t-1}$, generally denoted by $p(\cdot\,; \nu,\mu_t|\F_{t-1})$, whose dependence on a time-dependent quantity of interest $\mu_t$ and a distributional parameter $\nu$ are made explicit in the notation. Suppose $\mu_t\in S\subseteq\R$ and let $g:S\rightarrow\R$ be a twice differentiable, strictly monotonic one-to-one link function. The systematic component in a GLARMA$(p,q)$ model is specified through
\begin{equation}\label{model}
W_t = g(\mu_t) = \bs X_t'\bs\beta + Z_t,\quad\text{with }\quad\ Z_t:=\sum_{i=1}^p\phi_i(Z_{t-i}+\eps_{t-i})+\sum_{j=1}^q\theta_j\eps_{t-j},
\end{equation}
where $\eps_t:=\frac{Y_t-\mu_t}{V(\mu_t)^\lambda}$, for $\lambda\in(0,1]$, where $V:S\rightarrow (0,\infty)$ is the conditional variance of $Y_t$ given $\F_{t-1}$. The parameter $\lambda$ defines the scaling of the conditional residuals. Specifically, setting $\lambda = 1$ corresponds to the generalized residuals (often related to the score function), while $\lambda = 0.5$ yields the Pearson-type residuals. This choice is crucial as it determines how the magnitude of the prediction error updates the systematic component $W_t$.

In \eqref{model}, $\bs\beta:=(\beta_0,\cdots,\beta_r)'$ is a $(r+1)$-dimensional parameter associated with the covariates, $\bs\phi:=(\phi_1,\cdots,\phi_p)'$ and $\bs\theta:=(\theta_1,\cdots,\theta_q)'$ are the parameters associated to the AR and MA parts in $Z_t$, respectively. The term $Z_t$ acts as an observation-driven filter. The term $\eps_t$ represents the generalized residual that updates the systematic component, effectively driving the model's dynamics based on the one-step-ahead prediction error. The recursion in \eqref{model} requires initial values for the state and residuals, typically set as $Z_t=0$ and $\eps_t=0$ for $t \le 0$.

Initially developed for discrete distributions within the exponential family, such as Poisson, Binomial, and Negative Binomial \citep{Dunsmuir2016}, the GLARMA class has been widely used for count and binary time series. More recently, this framework was formally extended to accommodate continuous positive data through the incorporation of Gamma and Inverse Gaussian distributions \citep{Maia2026}, maintaining the original recursive filtering structure for the systematic component.

Due to the observation-driven nature of the GLARMA framework, the joint density of the observed series $\{Y_1, \dots, Y_n\}$ can be factorized as a product of conditional densities. Let $p(\cdot\,;\mu_t,\nu | \F_{t-1})$ denote the conditional probability density (or mass) function of $Y_t$ given its past history $\F_{t-1}$, where $\bs{\gamma} := (\nu,\bs \beta', \bs\phi', \bs\theta')'$ is the full vector of parameters. The conditional log-likelihood for a sample of size $n$ is given by
\begin{equation}\label{log_like_generic}
L(\bs{\gamma}) = \sum_{t=1}^{n} \ell_t(\boldsymbol{\bs{\gamma}}) = \sum_{t=1}^{n} \log \big(p(Y_t;\mu_t,\nu | \F_{t-1})\big).
\end{equation}
In this general formulation, the dependence of $p(\cdot)$ on the parameters $\bs{\gamma}$ occurs primarily through the time-varying quantity $\mu_t = g^{-1}(W_t)$, where $W_t$ follows the recursion defined in \eqref{model}. Numerical optimization methods, such as the Newton-Raphson algorithm or its variants, are typically employed to find the maximum likelihood estimate (MLE) $\hat{\bs{\gamma}}$. This requires the computation of the score vector $\mathbf{U}(\bs{\gamma}) = \partial L(\bs{\gamma}) / \partial \bs{\gamma}$, which, due to the recursive nature of $Z_t$, is evaluated using the chain rule ad
\begin{equation}\label{score_chain}
\frac{\partial \ell_t}{\partial \bs{\gamma}} = \frac{\partial \ell_t}{\partial \mu_t} \frac{\partial \mu_t}{\partial W_t} \frac{\partial W_t}{\partial \bs{\gamma}}.
\end{equation}
For further details on the frequentist estimation and the asymptotic properties of these estimators under different distributional assumptions, see \cite{Dunsmuir2016}. This conditional framework also provides a natural starting point for Bayesian inference, where the likelihood \eqref{log_like_generic} is combined with prior distributions to obtain the posterior density of the parameters.

\section{Bayesian inference} \label{Bayesian_section}
Let $\{Y_t\}_{t\in\Z}$ be a GLARMA$(p,q)$ model of interest with $r$-dimensional covariates $\{\bs X_t\}_{t\in\Z}$, $Y_1,\cdots,Y_n$ be a sample from it, and $\bs{\gamma} := (\nu,\bs \beta', \bs\phi', \bs\theta')'$. The conditional likelihood function is given by
\begin{equation*}
L(\bs\gamma|\F_{n})\propto\prod_{t=1}^n p(Y_t;\mu_t,\nu|\F_{t-1}). 
\end{equation*}
For the distributional parameter $\nu$, a compatible prior needs to be chosen. Usual weak/non-informative prior choices are 
\begin{equation*}
\pi_0(\nu)\sim
\begin{cases}
N(0,b),\ \text{for large} \ b,& \text{ if }\ \nu\in\R;\\
\mathrm{Gamma}(a,b),\ \text{with mean }\ a/b, \ \text{for }a\gg b, & \text{ if }\ \nu>0;\\
\mathrm{U}(a,b), &  \text{ if }\ a<\nu<b.\\
\end{cases}
\end{equation*}
The link function in \eqref{model} guarantees that $\mu_t\in S$ for all $\bs\beta\in\R^{r+1}$, $\bs\theta\in\R^q$ and $\bs\phi\in\R^p$, thus we can  assume independent multivariate normal priors of the form
\begin{equation}\label{mutla}
\bs\beta \sim N_{r+1}(\bs 0,\sigma_\beta^2I_{r+1}),\quad\bs\phi \sim N_p(\bs0,\sigma_\phi^2I_p),\quad \bs\theta\sim N_q(\bs0,\sigma_\theta^2I_q),
\end{equation}
where $N_s(\bs 0,K)$ denotes the $s$-variate normal distribution with mean $\bs0\in\R^s$ and variance-covariance matrix $K$, while $I_s$ denotes the $s\times s$ identity matrix. Alternatively, it may be more convenient to consider uniform distribution the AR and MA coefficients, such as $\phi_i\sim U(-1,1)$ and $\theta_i\sim U(-1,1)$, and also for parameter $\nu>0$, which can be prescribed using a $U(0,m)$ for some (large) $m>0$. These are all sensible choices for prior distribution, and using one instead of the other may depend on the particular the situation/application. 

In the absence of prior information for the variance hyper-parameters $\sigma_\beta^2$, $\sigma_\phi^2$, and $\sigma_\theta^2$, we assign large values to make the priors weakly informative. The posterior distribution is then obtained by combining the joint prior density $\pi_0(\bs\gamma)$ with the likelihood function:
\begin{equation} \label{post}
\pi(\bs\gamma|\F_{n})\propto L(\bs\gamma|\F_{n})\pi_0(\bs\gamma).
\end{equation}
As it is often the case, the joint posterior density \eqref{post} cannot be obtained in closed form. To sample from the posterior distribution we adopt a Monte Carlo scheme.

\section{Monte Carlo Study}
To evaluate the proposed Bayesian approach, we shall employ a Monte Carlo simulation study. We consider three fundamentally different GLARMA models to compared, chosen to showcase the flexibility of our approach. 
The first one is the Negative Binomial-GLARMA model for time series of counts, for which the random component $Y_t|\F_{t-1}$ follows a Negative Binomial distribution. The second one in the Beta-GLARMA model for double bounded time series, more specifically, taking values in $(0,1)$, for which $Y_t|\F_{t-1}$ follows a Beta distribution parameterized by its mean and a dispersion parameter $\nu$. The third model we consider is a model for positive time series: the Gamma-GLARMA model, for which for $Y_t|\F_{t-1}$ follows a Gamma distribution parameterized by its mean. 

In this numerical study, we focus on the GLARMA(1,0) model, guided by two main factors. First, pure autoregressive (AR) structures are heavily predominant in empirical applications, capturing the essential temporal structure observed in most real-world continuous and discrete phenomena. Second, from a numerical standpoint, MA components are inherently well-behaved and stable, whereas AR parameters close to the boundary of the stationarity domain present the most critical challenges for likelihood exploration. Therefore, constraining our analysis to the AR(1) structure allows for a rigorous stress test of the proposed MCMC framework under demanding dynamic regimes.

\subsection{First Scenario: the Negative Binomial-GLARMA}

To evaluate the performance and robustness of the proposed Bayesian framework in a discrete support context characterized by overdispersion under various temporal dynamics, we consider a Negative Binomial-GLARMA(1,0) model. The conditional distribution of the response variable $Y_t$, given the past history $\F_{t-1}$, is parameterized such that $Y_t | \F_{t-1} \sim \text{NB}(r, p_t)$, where $r > 0$ represents the dispersion (or size) parameter and $p_t = r / (r + \mu_t)$. Under this formulation, the conditional mean is given by $E(Y_t | \F_{t-1}) = \mu_t$ and the conditional variance is $\text{Var}(Y_t | \F_{t-1}) = \mu_t + \mu_t^2 / r$. The linear predictor incorporates a logarithmic link function, $\log(\mu_t) = X_t^\top \beta + Z_t$, ensuring the positivity of the conditional mean.

For this simulation scenario, the true structural parameter vector is specified as $\boldsymbol{\beta} = (\alpha, \beta_1, \beta_2)^\top = (1.0, 0.5, -0.2)^\top$ and the nuisance dispersion parameter is fixed at $r = 5.0$, introducing a moderate-to-high level of overdispersion into the simulated series. To analyze the estimator's ability to recover the true parameters under distinct dependence structures, we vary the autoregressive parameter across four regimes covering negative and positive persistence: $\phi \in \{-0.8, -0.3, 0.3, 0.8\}$. Bayesian estimation is conducted by specifying weakly informative prior distributions to ensure numerical stability while letting the data dominate the posterior inference. Specifically, we assign independent standard normal priors for the regression coefficients, $\alpha, \beta_1, \beta_2 \sim N(0, 1)$, a continuous uniform prior over the invertibility region for the autoregressive parameter, $\phi \sim U(-1, 1)$, and a flat uniform prior for the dispersion parameter, $r \sim U(0, 50)$.

The results of the MCMC simulation across the 50 Monte Carlo replicas for each regime of $\phi$ are summarized in Table~\ref{tab:sim_nb_results}. In general, chain diagnostics indicated excellent mixing properties and instantaneous convergence for all scenarios, with potential scale reduction factors (Gelman-Rubin diagnostic) strictly close to $1.0$ and negligible Monte Carlo standard errors. Due to rapid stabilization, we experimented with applying burn-in periods and thinning intervals during the pilot stages of the simulation study. However, since these adjustments had no significative effect on the quality of the posterior distributions or on reducing autocorrelation within the chains, we retained the full length of iterations for all generated chains without burn-in or thinning.

\begin{table}[ht]
\centering
\setlength{\tabcolsep}{7pt} 
\renewcommand{\arraystretch}{1.3}
\caption{Median parameter estimates (Est.) and coefficients of variation (CV) for Negative Binomial-GLARMA$(1,0)$ models under weakly informative priors, averaged over $50$ replications with $n=1000$ and chain length $6000$. Results are shown for $\phi\in\{-0.8,-0.3,0.3,0.8\}$, $\alpha=1$, $\beta_1=0.5$, $\beta_2=-0.2$, and $r=5$.}\vspace{0.3cm}
\label{tab:sim_nb_results}
\begin{tabular}{c|cc|cc|cc|cc}
\multicolumn{1}{c}{} & 
     \multicolumn{2}{c}{$\phi=-0.8$} & \multicolumn{2}{c}{$\phi=-0.3$} & \multicolumn{2}{c}{$\phi=0.3$} & \multicolumn{2}{c}{$\phi=0.8$}\\ 
    \hline
\multicolumn{1}{c}{}    &  Est. & CV &  Est. & CV &  Est. & CV &  Est. & CV  \\ 
 \hline 
$\hat\alpha$   & 1.001 & 1.110\%  & 1.004  & 1.800\%  & 0.998  & 3.430\%  & 0.973  & 11.580\% \\
$\hat\beta_1$  & 0.501 & 3.320\%  & 0.493  & 4.810\%  & 0.500  & 4.990\%  & 0.501  & 1.530\%  \\
$\hat\beta_2$  & -0.201& 6.660\%  & -0.201 & 9.950\%  & -0.195 & 10.880\% & -0.200 & 3.260\%  \\
$\hat\phi$     & -0.797& 1.760\%  & -0.289 & 9.310\%  & 0.302  & 7.100\%  & 0.801  & 1.530\%  \\ 
$\hat r$       & 5.112 & 8.600\%  & 5.328  & 22.140\% & 5.224  & 15.920\% & 4.980  & 6.010\%  \\ 
 \hline
\end{tabular}
\end{table}
Table~\ref{tab:sim_nb_results} reveals that the proposed Bayesian approach achieves high accuracy and precision in retrieving the true parameter values across all evaluated persistence levels under a Negative Binomial context. The structural regression coefficients ($\hat{\bs\beta}$) are successfully recovered, displaying small deviations from their nominal values and remarkably low Coefficient of Variation (CV) values. 

Interestingly, under a strong positive autocorrelation regime ($\phi = 0.8$), the baseline level estimator $\hat\alpha$ displays a higher relative variability ($\text{CV} = 11.580\%$) paired with a minor underestimation ($\text{Est.} = 0.973$). This empirical finding can be directly connected to the inherent numerical challenges of parameter estimation in observation-driven frameworks near the boundary of the stationarity domain. As highlighted by \cite{Dunsmuir2016}, when autoregressive parameters approach the edge of the unit circle, classical frequentist numerical optimization based on gradient methods often encounters severe convergence difficulties and parameter instability, driven by the persistent and long-lasting impact of the filter's initial conditions. Furthermore, as documented by \citet{Camara2025} within a bootstrap inference context for count time series, such strong dynamic persistence and the underlying dependence structure can heavily dictate the stability and finite-sample behavior of GLARMA estimators. In our current Bayesian setup, however, while this high-persistence regime induces a marginal inflation in the intercept's relative variability, the simulation remains entirely stable across all replications, completely bypassing numerical breakdowns or MCMC non-convergence issues. Conversely, the covariate effects $\hat\beta_1$ and $\hat\beta_2$ under this exact same regime achieve their maximum relative precision, yielding the lowest CV values observed in the entire experiment ($1.530\%$ and $3.260\%$, respectively). This empirical evidence suggests that the MCMC framework via \texttt{Nimble} offers a highly reliable alternative, preserving the structural estimation and interpretation of regression slopes even under challenging, high-frequency short-term memory setups.

The autoregressive parameter $\hat\phi$, which regulates the recursive feedback of past prediction errors through the Pearson innovations, is recovered with remarkable accuracy across all dynamic setups. The median estimates for $\hat\phi$ remain exceptionally close to their nominal values, ranging from $-0.797$ (when $\phi = -0.8$) to $0.801$ (when $\phi = 0.8$). Furthermore, its relative variability is tightly controlled, with the CV reaching a minimum of $1.530\%$ under strong positive persistence and a maximum of $9.310\%$ under mild negative dependence ($\phi = -0.3$). This indicates that the MCMC algorithm successfully isolates the short-term serial dependence from both the overdispersion effect and the covariate effects, regardless of whether the process displays alternating dynamics ($\phi < 0$) or high persistence ($\phi > 0$).

Finally, the nuisance parameter $\hat{r}$ displays the characteristic behavior expected for dispersion parameters in generalized linear time series frameworks. For the mild persistence regimes ($\phi = -0.3$ and $\phi = 0.3$), the flatter likelihood surface typical of dispersion parameters under finite sample sizes ($n = 1000$) results in a moderate overestimation ($\text{Est.} = 5.328$ and $5.224$, respectively) accompanied by higher relative variability ($\text{CV} = 22.140\%$ and $15.920\%$). However, a striking and highly desirable result is achieved under strong temporal dynamics. For both $\phi = -0.8$ and $\phi = 0.8$, the estimation of $\hat{r}$ becomes significantly more robust and accurate, with the estimates converging remarkably close to the target ($\text{Est.} = 5.112$ and $4.980$) and the relative variability dropping sharply ($\text{CV} = 8.600\%$ and $6.010\%$). This pattern suggests that when the underlying time series features prominent, high-frequency oscillations or long-lasting memory structures, the recursive filtering mechanism embedded within the GLARMA framework accumulates conditional information more effectively, stabilizing the conditional variance and considerably enhancing the algorithm's capability to identify and recover the overdispersion parameter.
\subsection{Second Scenario: the Beta-GLARMA}\label{sec:beta}

In this section, we consider the case where the sample is generated from a Beta-GLARMA$(1,0)$ model. We adopt the beta distribution parameterized by its mean, with density given by
\[
p(y;\mu_t,\nu) = \frac{\Gamma(\nu)}{\Gamma(\nu\mu_t)\Gamma\big(\nu(1-\mu_t)\big)}y^{\nu\mu_t-1}(1-y)^{\nu(1-\mu_t)-1}I(0<y<1),
\]
for $\nu>0$, whereas the systematic component follows \eqref{model}. We consider Beta-GLARMA models with $|\phi|\in\{0.3,0.8\}$, intercept $\alpha = 1$, and $\nu=4$. In this case, the variance function is $\text{Var}(Y_t | \F_{t-1}) =\frac{\mu_t(1-\mu_t)}{1+\nu}$. The sample size was fixed at $n=1000$. For each scenario, a single Markov chain of length $6000$ was generated using \texttt{Nimble}.

Estimation of Beta-GLARMA$(1,0)$ models is challenging under both frequentist and Bayesian approaches. In a pilot simulation, we found that the parameters $\nu$ and $\phi$ were highly sensitive to the choice of prior distribution. In the pilot study, we used the priors $\nu\sim U(0,10)$, $\phi\sim U(-0.9,0.9)$, and $\alpha\sim N(0,1)$. We also experimented with Gamma priors for $\nu$. All configurations led to poor estimation of $\nu$ and $\phi$. To mitigate this, we adopted slightly more informative priors: $\nu\sim U(3,7)$, and $\phi$ following a uniform distribution centered at the true value of $\phi$ with a half-width of $0.3$. This configuration considerably improved the original results. 

In practice, however, the true value of the parameter is not known to apply these informative priors. In these situations, historical data or results from other similar experiments can be used to refine the prior specification. Nevertheless, the strong dependence of the estimates on the prior choice observed in the pilot study raises important questions about the robustness of the inference to prior misspecification. This issue is particularly relevant for $\phi$, given its direct role in the autoregressive structure of the model. To investigate this matter further, we conduct a systematic sensitivity analysis in the following section, where we examine the behavior of the model under different prior specifications for $\phi$, including correctly specified, incorrectly specified, and non-informative cases.

Table~\ref{tab:beta} presents the results in this more informative scenario. There we report, for each parameter, the median estimate averaged over the $50$ replications (column Est.) and the coefficient of variation (CV), which facilitates comparison of variability across parameters with different scales. Although chain diagnostics were all positive, we also experimented with applying burn-in and/or thinning, but these had no effect on the quality of the results; therefore, we retained all chains without burn-in or thinning. 

The results show that $\phi$ is consistently well estimated, even in the more challenging scenarios with $|\phi|=0.8$. The estimates of $\nu$ and $\alpha$ appear to be influenced by the magnitude of $\phi$. For $\alpha$, under smaller values of $\phi$, the estimates are very close to the nominal value $\alpha=1$, whereas some bias is observable under stronger dependence. Parameter $\nu$ is the most challenging to estimate, exhibiting considerable bias even when $|\phi|=0.3$. Compared with the related results presented for $\beta$ARMA models in \cite{Aline}, where $\nu$ was also the most difficult parameter to estimate, in the Beta-GLARMA there is an extra layer of complexity in the model: the error term in the systematic component explicitly depends on $\nu$. This implies a direct dependence between $\nu$ and the parameters in the systematic component, absent in the $\beta$ARMA case. This complexity may account for the added difficulty in estimating $\nu$ in the present case.

The results obtained under the informative priors, while encouraging, must be interpreted with caution given their reliance on prior information that would be unavailable in genuine applications. This motivates the sensitivity analysis that follows, where we systematically evaluate how the estimates respond to different prior choices, thereby providing guidance on the extent to which the results can be trusted under realistic data-analytic conditions.
\begin{table}[ht]
\centering
\setlength{\tabcolsep}{7pt} 
\renewcommand{\arraystretch}{1.3}
\caption{Median parameter estimates (Est.) and coefficients of variation (CV) for Beta-GLARMA$(1,0)$ models under informative priors, averaged over $50$ replications with $n=1000$ and chain length $6000$. Results are shown for $|\phi|\in\{0.3,0.8\}$, $\alpha=1$, and $\nu=4$.}\vspace{0.3cm}
\label{tab:beta}
\begin{tabular}{c|cc|cc|cc|cc}
\multicolumn{1}{c}{} & 
     \multicolumn{2}{c}{$\phi=-0.8$} & \multicolumn{2}{c}{$\phi=-0.3$} & \multicolumn{2}{c}{$\phi=0.3$} & \multicolumn{2}{c}{$\phi=0.8$}\\ 
    \hline
\multicolumn{1}{c}{}    &  Est. & CV &  Est. & CV &  Est. & CV &  Est. & CV  \\ 
 \hline 
$\hat\alpha$ & 1.185 & 3.273\% & 0.988 & 2.479\% & 0.978 & 5.150\% & 1.143 & 11.111\%\\
 $\hat\phi$ & -0.799 & 0.917\% & -0.300 & 2.121\% & 0.299 & 2.528\% & 0.799 & 1.133\%\\ 
 $\hat\nu$ & 3.003 & 0.014\% & 3.604 & 4.366\% & 3.552 & 5.214\% & 3.003 & 0.015\%\\ 
 \hline
\end{tabular}
\end{table}
\FloatBarrier

\subsubsection{Sensitivity Analysis}

We conduct a sensitivity analysis to assess the influence of prior specification on the estimation of parameter $\phi$ in Beta-GLARMA models. We consider models with true parameter values $\phi \in \{0.3, 0.6\}$, $\alpha = 1$, and $\nu = 4$. For each value of $\phi$, we evaluate three classes of priors: (i) informative and correctly specified, where the prior distribution is a relatively narrow interval containing the true $\phi$; (ii) informative and incorrectly specified, where the prior is narrow but does not contain the true $\phi$; and (iii) non-informative, represented by a wide uniform prior. For $\phi = 0.3$, the priors examined are $U(0.1,0.4)$ and $U(0,0.5)$ (correctly specified), $U(-0.1,0.2)$ and $U(0.5,0.7)$ (incorrectly specified), and $U(-0.9,0.9)$ (non-informative). An analogous set of priors is used for $\phi = 0.6$: $U(0.4,0.8)$ and $U(0.3,0.9)$ (correctly specified), $U(0.2,0.5)$ and $U(0.7,0.9)$ (incorrectly specified), and $U(-0.9,0.9)$ (non-informative). For the remaining parameters, we adopt weakly informative priors: $\alpha \sim N(0,1)$ and $\nu \sim U(2,10)$. Observe that the prior for $\nu$ is less informative than the one used in Section~\ref{sec:beta}. Table~\ref{tab:sens} presents the median posterior estimates and coefficients of variation (CV) for each parameter under the different prior specifications.

\begin{table}[ht]
\centering
\small
\caption{Median posterior estimates and coefficients of variation (CV) for parameters $\alpha$, $\phi$, and $\nu$ under different prior specifications for $\phi$. Results are shown for true values $\phi = 0.3$ and $\phi = 0.6$, with informative correctly specified, informative incorrectly specified, and non-informative priors.}\vspace{.3cm}
\label{tab:sens}
\begin{tabular}{c|cc|cc|cc|cc|cc}\hline
 \multicolumn{11}{c}{$\phi=0.3$}\\ 
 \hline
\multicolumn{1}{c|}{Prior}& \multicolumn{2}{c|}{$U(0.1,0.4)$} & \multicolumn{2}{c|}{$U(0,0.5)$} & \multicolumn{2}{c|}{$U(-0.9,0.9)$} & \multicolumn{2}{c|}{$U(-0.1,0.2)$} & \multicolumn{2}{c}{$U(0.5,0.7)$} \\ \hline
\multicolumn{1}{c|}{Param.}  &  Est. & CV &  Est. & CV  &  Est. & CV &  Est. & CV & Est. & CV  \\ 
\hline 
$\alpha$  &0.982 & 4.305\% & 0.995 & 4.458\% & 0.976 & 3.642\% & 0.982 & 4.629\% & 0.983 & 4.168\%\\ 
 $\phi$  &0.250 & 1.754\% & 0.250 & 2.246\% & 0.005 & 470.6\% & 0.050 & 7.028\% & 0.600 & 0.565\%\\ 
 $\nu$  &3.630 & 3.773\% & 3.603 & 4.467\% & 3.563 & 4.393\% & 3.542 & 5.278\% & 3.591 & 4.008\%\\ 
 \hline
 \multicolumn{11}{c}{$\phi=0.6$}\\ 
 \hline\multicolumn{1}{c|}{Prior}& \multicolumn{2}{c|}{$U(0.4,0.8)$} & \multicolumn{2}{c|}{$U(0.3,0.9)$} & \multicolumn{2}{c|}{$U(-0.9,0.9)$} & \multicolumn{2}{c|}{$U(0.2,0.5)$} & \multicolumn{2}{c}{$U(0.7,0.9)$} \\ \hline
\multicolumn{1}{c|}{Param.}  &  Est. & CV &  Est. & CV  &  Est. & CV &  Est. & CV & Est. & CV  \\ 
\hline 
$\alpha$  &0.894 & 7.255\% & 0.926 & 8.154\% & 0.904 & 7.680\% & 0.895 & 8.301\% & 0.915 & 9.557\%\\ 
 $\phi$  &0.600 & 0.922\% & 0.601 & 1.104\% & 0.003 & 1011.1\% & 0.350 & 1.076\% & 0.800 & 0.324\%\\ 
 $\nu$  &2.420 & 6.102\% & 2.432 & 5.087\% & 2.385 & 5.175\% & 2.395 & 6.558\% & 2.431 & 7.252\%\\ 
 \hline
\end{tabular}
\end{table}

The sensitivity analysis reveals several important patterns. For both $\phi = 0.3$ and $\phi = 0.6$, the correctly specified informative priors yield median estimates of $\phi$ that closely approximate the true values, with remarkably low coefficients of variation (CV $< 2.3\%$ for $\phi = 0.3$ and CV $< 1.2\%$ for $\phi = 0.6$). This indicates that when prior information is accurate, the model produces highly stable and precise estimates. Conversely, the non-informative prior $U(-0.9,0.9)$ leads to severely biased estimates of $\phi$ (0.005 and 0.003 for $\phi = 0.3$ and $0.6$, respectively) with extremely high CVs (470.6\% and 1011.1\%), suggesting that the data alone may not sufficiently identify $\phi$ without appropriate prior information.

When the prior is misspecified, the estimates of $\phi$ are pulled toward the prior mean rather than the true value. For $\phi = 0.3$, the incorrectly specified prior $U(-0.1,0.2)$ yields a median estimate of 0.050, while $U(0.5,0.7)$ produces an estimate of 0.600 -- both reflecting the prior's location. Similarly, for $\phi = 0.6$, the misspecified priors $U(0.2,0.5)$ and $U(0.7,0.9)$ result in estimates of 0.350 and 0.800, respectively. Notably, the CVs remain relatively low under misspecification, indicating that the model is precise but systematically biased when the prior is incorrect.

Estimates of $\alpha$ and $\nu$ remain relatively stable across all prior specifications for $\phi$, with $\alpha$ centered near the true value of 1 and $\nu$ slightly underestimated (ranging from 2.385 to 3.630, compared to the true value of 4). This suggests that the prior for $\phi$ has limited impact on the estimation of the other model parameters, and any bias in $\nu$ may be attributable to the weakly informative prior $U(2,10)$ or local model identifiability issues.

Overall, these findings highlight the critical role of prior specification for $\phi$ in Beta-GLARMA models. While informative priors can substantially improve estimation precision when correctly specified, they can also introduce significant bias when misspecified. The non-informative prior performs poorly, suggesting that prior information about $\phi$ is essential for reliable inference in this modeling framework.
\subsection{Third Scenario: the Gamma-GLARMA}

The MCMC simulation results across the 50 Monte Carlo replicas for each regime of $\phi$ under the continuous Gamma-GLARMA(1,0) specification are summarized in Table~\ref{tab:gamma}. For this scenario, Bayesian estimation is conducted by assigning weakly informative prior distributions to ensure numerical stability: independent standard normal priors for the regression coefficients, $\alpha, \beta_1, \beta_2 \sim N(0, 1)$, a continuous uniform prior over the stationary domain for the persistence parameter, $\phi \sim U(-1, 1)$, and a flat uniform prior for the shape parameter, $\nu \sim U(0, 50)$. Overall, the parameters of the linear predictor ($\hat\alpha, \hat\beta_1, \hat\beta_2$) are accurately recovered, presenting small deviations from their true values and highly controlled relative variability. Interestingly, under the extreme positive persistence regime ($\phi = 0.8$), the baseline level estimator $\hat\alpha$ displays an increased coefficient of variation ($\text{CV} = 13.360\%$) and a slight underestimation ($\text{Est.} = 0.983$). Conversely, an even more pronounced inflation in relative variability is observed under the strong alternating dynamic setup ($\phi = -0.8$), where $\hat\alpha$ yields a $\text{CV}$ of $21.130\%$ and a moderate overestimation ($\text{Est.} = 1.054$). This pattern shows that the intercept estimator becomes less precise under extreme positive or negative autocorrelation regimes. Reassuringly, the estimation of the regression slopes remains remarkably robust; the slope estimators $\hat\beta_1$ and $\hat\beta_2$ maintain highly stable distributions across all scenarios, with their CVs never exceeding $7.510\%$ and $19.110\%$, respectively, successfully isolating the covariate impacts from the temporal structure.

\begin{table}[ht]
\centering
\setlength{\tabcolsep}{7pt} 
\renewcommand{\arraystretch}{1.3}
\caption{Median parameter estimates (Est.) and coefficients of variation (CV) for Gamma-GLARMA$(1,0)$ models under weakly informative priors, averaged over $50$ replications with $n=1000$ and chain length $6000$. Results are shown for $\phi\in\{-0.8,-0.3,0.3,0.8\}$, $\alpha=1$, $\beta_1=0.5$, $\beta_2=-0.2$, and $\nu=0.8$.}\vspace{0.3cm}
\label{tab:gamma}
\begin{tabular}{c|cc|cc|cc|cc}
\multicolumn{1}{c}{} & 
     \multicolumn{2}{c}{$\phi=-0.8$} & \multicolumn{2}{c}{$\phi=-0.3$} & \multicolumn{2}{c}{$\phi=0.3$} & \multicolumn{2}{c}{$\phi=0.8$}\\ 
    \hline
\multicolumn{1}{c}{}    &  Est. & CV &  Est. & CV &  Est. & CV &  Est. & CV  \\ 
 \hline 
$\hat\alpha$   & 1.054 & 21.130\% & 1.002  & 3.230\%  & 1.000  & 5.170\%  & 0.983  & 13.360\% \\
$\hat\beta_1$  & 0.489 & 7.510\%  & 0.501  & 5.700\%  & 0.504  & 6.080\%  & 0.502  & 3.870\%  \\
$\hat\beta_2$  & -0.191& 18.590\% & -0.198 & 19.110\% & -0.199 & 15.900\% & -0.200 & 10.510\% \\
$\hat\phi$     & -0.753& 30.060\% & -0.302 & 11.330\% & 0.301  & 11.600\% & 0.797  & 3.100\%  \\ 
$\hat\nu$      & 0.786 & 12.380\% & 0.804  & 3.710\%  & 0.801  & 3.640\%  & 0.799  & 1.800\%  \\ 
 \hline
\end{tabular}
\end{table}

The autoregressive parameter $\hat\phi$, which governs the dynamic error propagation via the Pearson innovations, displays a highly asymmetric precision pattern between the negative and positive regimes. For the mild persistence scenarios ($\phi = -0.3$ and $\phi = 0.3$), the median estimates are precise, centering at $-0.302$ ($\text{CV} = 11.330\%$) and $0.301$ ($\text{CV} = 11.600\%$). However, when evaluating the boundaries of the stationarity domain, the algorithm's performance provides a compelling discussion. Under $\phi = 0.8$, where the parameter lies in close proximity to the unit circle boundary, the recovery of the dynamic structure is exceptionally precise, with the estimate reaching $0.797$ and achieving its lowest relative variability ($\text{CV} = 3.100\%$). Notably, these results are obtained using the specified weakly informative setup, demonstrating that the MCMC algorithm converges near the boundary without requiring strongly restrictive or tightly informative prior distributions. In contrast, under the high-frequency alternating regime ($\phi = -0.8$), the estimator shows a mild attenuation effect ($\text{Est.} = -0.753$) accompanied by a sharp inflation in uncertainty ($\text{CV} = 30.060\%$). This boundary behavior can be evaluated in light of the numerical challenges documented by \citet{Dunsmuir2016}, where parameters close to the edge of the admissible space can trigger instabilities due to the lingering impact of the recursive filter's initial states. Nonetheless, our Bayesian framework demonstrates stability, converging successfully across all iterations without experiencing numerical breakdowns or algorithmic failures even under the most demanding $\phi = -0.8$ setup.

Finally, the Gamma shape parameter $\hat\nu$ is accurately recovered across all setups. In the mild persistence cases ($\phi = -0.3$ and $\phi = 0.3$), the Bayesian approach yields precise estimates of $0.804$ ($\text{CV} = 3.710\%$) and $0.801$ ($\text{CV} = 3.640\%$), respectively. Crucially, stronger temporal dynamics further improve estimation efficiency. Under strong positive correlation ($\phi = 0.8$), $\hat\nu$ converges almost perfectly to the target ($\text{Est.} = 0.799$) with its lowest relative variability ($\text{CV} = 1.800\%$), while a similar stabilization occurs under $\phi = -0.8$ ($\text{Est.} = 0.786$, $\text{CV} = 12.380\%$). These results indicate that higher serial dependence enhances the recursive filter's ability to update and stabilize the conditional variance, enabling a highly reliable estimation of the model's dispersion parameter via MCMC.

\section{Empirical Analysis}
In this section we present two empirical analysis to showcase the usefulness of the proposed Bayesian approach.

\subsection{Nuclear Energy Net Generation - United States}

\begin{figure}[!ht]
\centering
\includegraphics[width=0.65\textwidth]{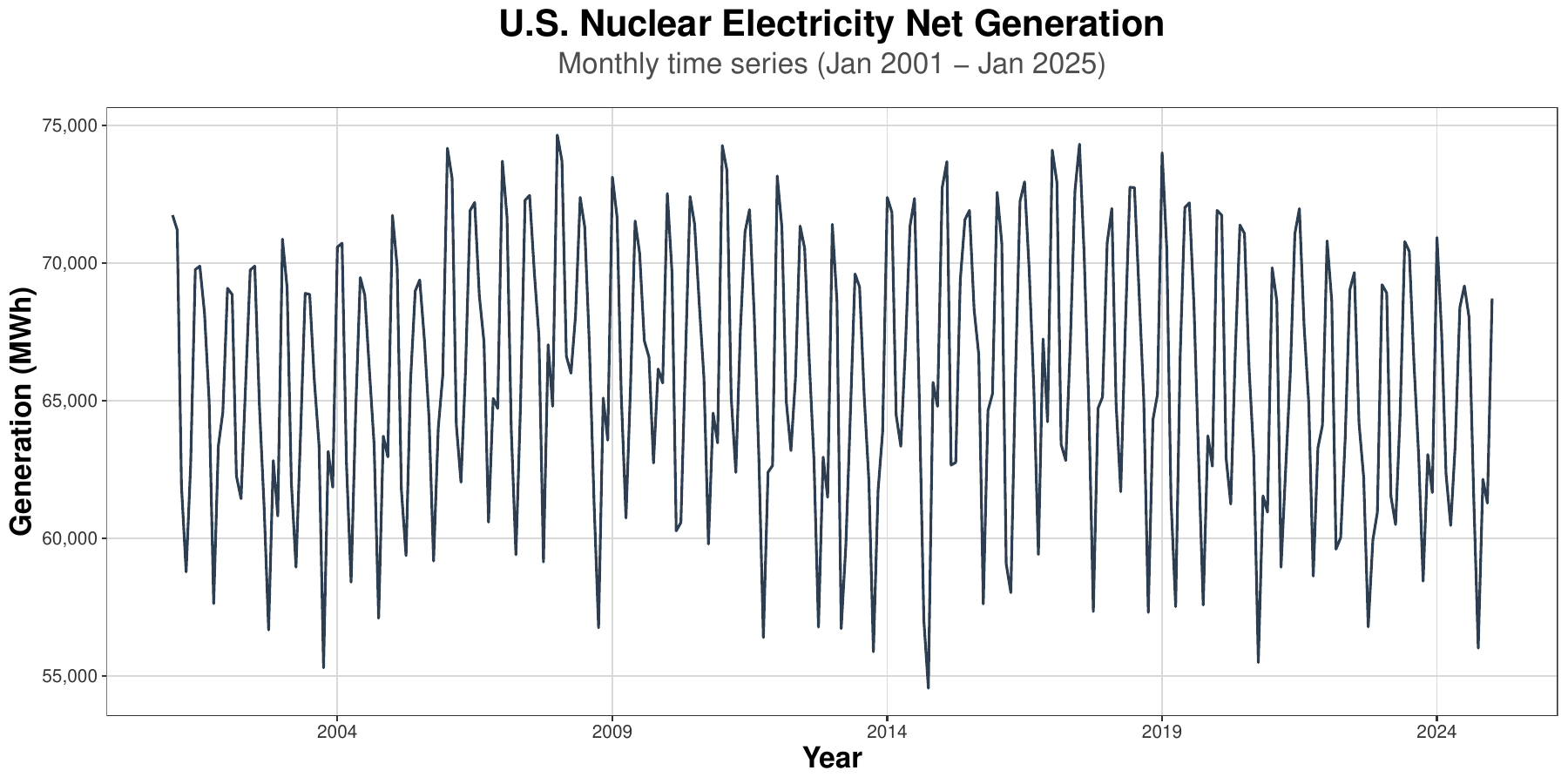}
\includegraphics[width=0.65\textwidth]{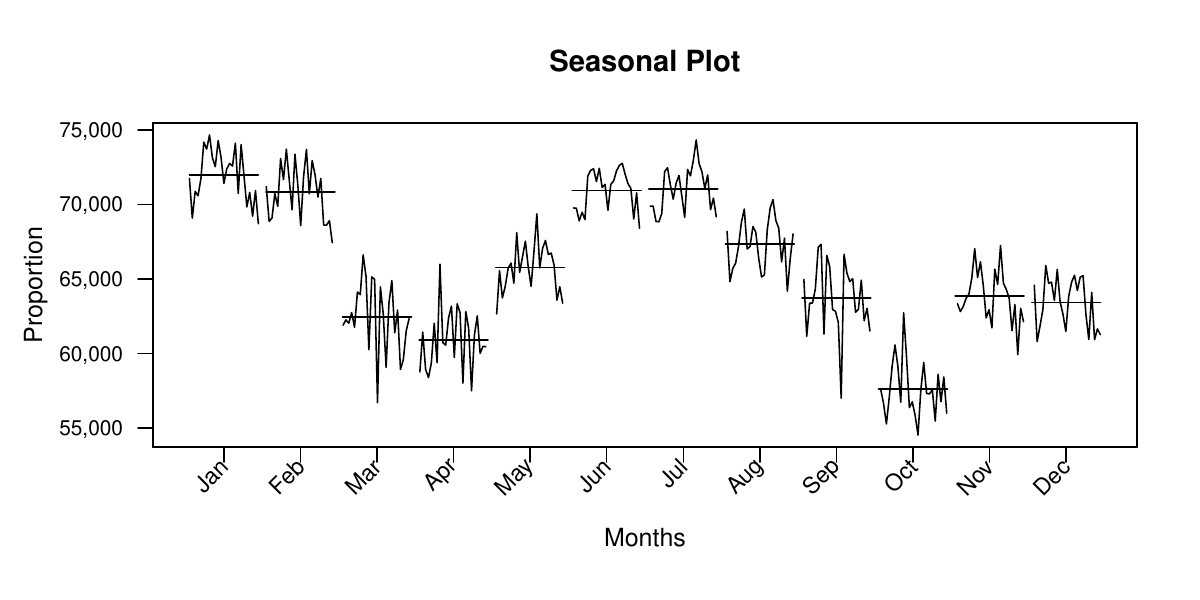}
\caption{Time series plot of the data showing a distinct seasonal pattern along with the seasonal plot}
\label{fig:season}
\end{figure}

In this section we employ the proposed Bayesian approach for GLARMA models to analyze the monthly electricity generated by nuclear plants for all sectors in the United States. The raw data are freely available from the U.S. Energy Information Administration's (EIA) website\footnote{{\color{blue} \href{https://www.eia.gov/electricity/data/browser/\#/topic/}{www.eia.gov/electricity/data/browser/\#/topic/}}, retrieved 11/26/2025} and contain monthly net energy generated for all sectors of the United States, including coal, petroleum liquids, petroleum coke, natural gas, other gases, nuclear, conventional hydroelectric, other renewables (total), hydro-electric pumped storage and other. We focus on the net electricity generated by nuclear plants in the US from January 2001 to January 2025, yielding a time series with $n=289$ observations. The time series plot along with a seasonal plot is presented in Figure~\ref{fig:season}. 

The seasonal plot reveals a clear and stable periodic pattern in the monthly nuclear electricity net generation. The most visible feature is a pronounced semi-annual cycle, with lower generation levels typically observed in the spring (April--May) and autumn (September--October) months, alongside a recurrent annual rhythm that reflects the broader seasonal demand cycle. These two periodicities, at lags 6 and 12 months, are not harmonics of each other, as their presence arises from a combination of operational and demand-driven factors: nuclear power plants generally operate at full capacity during peak demand periods (summer and winter), while scheduled refueling and maintenance outages are deliberately concentrated in the lower-demand transition seasons (spring and autumn). Consequently, the alternating pattern of high and low production every six months, superimposed on the full annual cycle, stems directly from these strategic operational decisions.

Given the strictly positive nature of the data, we employ a Bayesian Gamma-GLARMA approach. The steps used to obtain the final model are as follows. To model seasonality, we adopt a harmonic analysis strategy, including combinations of sines and cosines at different frequencies in the systematic component \eqref{model}. We then select the AR and MA components to be included. We use the following priors: $\alpha\sim N(0,25)$; $\beta_i\sim N(0,1)$; $\theta_i,\phi_i\sim U(-1,1)$ and $\nu\sim U(0,100)$. 

For each candidate model, we generated a single Markov chain of 50,000 iterations, discarding the first 5,000 as burn-in. To reduce storage and autocorrelation, we applied a thinning factor of 10, keeping every 10th observation. This yielded a final chain of 4,500 draws for inference.

Convergence was assessed using the Geweke and Heidelberger--Welch (Heidel) diagnostic tests, implemented via \texttt{geweke.diag} and \texttt{heidel.diag} from the \texttt{coda} package \citep{coda}, with default settings. We further computed the effective sample size (ESS) using \texttt{effectiveSize} also from \texttt{coda} and examined HPD and quantile-based credible intervals (CI). HPD CIs were constructed using the function \texttt{emp.hpd} from the \texttt{TeachingDemos} package \citep{hpd}, whereas the quantile-based ones were obtained using the respective quantiles from the posterior sample, as computed by the function \texttt{quantile} from the \texttt{stats} package, with default configurations. A model is considered suitable for residual analysis if all diagnostics indicate convergence, the ESS exceeds 1,000 for all parameters, and both HPD and quantile-based CIs exclude zero.

For residual analysis, we take the posterior median of each parameter as a point estimator and recursively reconstruct $\hat\mu_1,\dots,\hat\mu_n$ using \eqref{model}. We then compute Pearson residuals $\hat r_t^p := \sqrt{\hat\nu}(y_t-\hat\mu_t)/\hat\mu_t$. If the model is well specified, we expect the Pearson residuals to be a martingale difference sequence (MDS) with respect to the process history. To test for MDS, we apply the Domínguez--Lobato test \citep{dominguez}, which is robust against nonlinear and higher-order dependence. The test is implemented in the \texttt{BTSR} package \citep{BTSR}, considering lag 1 and with $p$-values calculated using 500 bootstrap samples. All tests were performed at the 5\% significance level. We consider the model fit satisfactory only when the null hypothesis of MDS is not rejected.

We sought the most parsimonious model that satisfies all the criteria presented above. We initially included sine and cosine covariates for frequencies 6 and 12, and, after further exploration, discovered that frequency 108 was also important. The specific AR and MA components were selected heuristically following a Box--Jenkins-like approach based on the Pearson residuals. Obtaining a final model required several attempts, in which we undertook the intermediate steps described above, involving the inclusion and exclusion of AR and MA terms, and proceeding with goodness-of-fit and residual analyses for every candidate model.

\subsection{Results}

After a number of trials, we arrived at a Gamma-GLARMA model with 5 covariates related to the seasonal component and a single autoregressive component of order 12 -- other models including lower-order autoregressive terms failed to satisfy the requirements presented in the previous section. The final fitted Gamma-GLARMA model, according to our criteria, is as follows:
\begin{equation*}
\log(\mu_t) = \alpha + \beta_1\sin\bigg(\frac{2\pi t}{6}\bigg)+\beta_2\cos\bigg(\frac{2\pi t}{6}\bigg)+\beta_3\sin\bigg(\frac{2\pi t}{12}\bigg)+\beta_4\cos\bigg(\frac{2\pi t}{12}\bigg)+\beta_5\sin\bigg(\frac{2\pi t}{108}\bigg)+Z_t
\end{equation*}
with $Z_t=\phi_{12}(Z_{t-12}+e_{t-12})$. The inclusion of the 108-month (9-year) cycle was not visually apparent from the exploratory analysis but emerged as a significant component during the modeling process. This longer cycle captures the slower, structural evolution of the nuclear fleet's generation capacity over time, reflecting factors such as plant retirements, life extensions, and the gradual commissioning of new units. Its presence is sometimes mistakenly confounded with long-range dependence in the first difference of the series \citep{Barros2013}. 

Table~\ref{tab:gfit} presents the fitted model results and convergence diagnostics. Point estimates are taken as the posterior median. Notably, the posterior mean and median agree to at least four decimal places for most parameters, with minor discrepancies only for $\beta_5$ ($-0.0169$ vs.~$-0.0168$) and $\nu$ ($770.746$ vs.~$771.792$). Regarding convergence, the Geweke, Heidel, and Ljung--Box tests yield $p$-values above $0.1322$ in all cases, providing no evidence against convergence. The effective sample sizes (all $>3979$) indicate excellent mixing and negligible autocorrelation. The reported HPD credible intervals exclude zero for every parameter, suggesting that all effects are statistically meaningful. This finding is corroborated by quantile-based credible intervals (not shown), which also do not contain zero. Overall, the diagnostics suggest that the chains have reached stationarity, exhibit high mixing, and display low to negligible serial dependence.

\begin{remark}
We found that the difference between using the whole chain or the burned-in and thinned one does not influence the results when the chain diagnostics suggest convergence. In these cases, using the mean or median also made little to no difference. In general, we observe differences in the 4th decimal place or smaller for all parameters except $\nu$, whose scale is larger. In any case, we decided to apply a burn-in and thinning in order to streamline the procedure.
\end{remark}

\begin{table}[ht]
\caption{Fitted model results and convergence diagnostics. For each parameter, we report the point estimate (posterior median), $p$-values for the Geweke, Heidel, and Ljung--Box tests, effective sample size (ESS), and the lower and upper bounds of the HPD credible intervals (CI inf and CI sup).}\label{tab:gfit}\vspace{.3cm}
\centering
\begin{tabular}{c|r|ccc|c|rr}
Par. &   Median  &    Geweke    &     Heidel    & Ljung-Box &  ESS & CI inf. & CI sup. \\
\hline
$\alpha$  &   11.095  &    0.5205    &     0.1322    &  0.1844   &  4376  & 11.090 & 11.097 \\
$\beta_1$ &   0.0729  &    0.7555    &     0.2654    &  0.7682   &  4500  & 0.0639  & 0.0772 \\
$\beta_2$ &   0.0411  &    0.1935    &     0.7698    &  0.9765   &  4500  & 0.0321  & 0.0456 \\
$\beta_3$ &   0.0161  &    0.3452    &     0.2230    &  0.3435   &  4291  & 0.0082  & 0.0197 \\ 
$\beta_4$ &  -0.0212  &    0.3008    &     0.7589    &  0.7972   &  4500  & -0.0288 & -0.0175 \\
$\beta_5$ &  -0.0169  &    0.9426    &     0.1567    &  0.2940   &  3979  & -0.0264 & -0.0127 \\
$\nu$     &   770.75  &    0.5558    &     0.7196    &  0.4019   &  4071  & 559.35  & 880.18 \\
$\phi_{12}$    &  -0.0124  &    0.3812    &     0.1525    &  0.2857   &  4189  & -0.0225 & -0.0084 \\
\hline
\end{tabular}
\end{table}

For the residual analysis, the Pearson residuals do not reject the MDS null hypothesis, with $p$-values of 0.050 (Cramér--von Mises) and 0.056 (Kolmogorov--Smirnov).

The fitted model provides valuable insights into the operational dynamics of the U.S. nuclear power fleet. The seasonal components confirm the expected semi-annual pattern of lower generation during spring and autumn transition months, reflecting planned maintenance and refueling outages that are deliberately scheduled outside peak demand periods. More importantly, the identification of a statistically significant 108-month (9-year) cycle, with a credible interval excluding zero ($-0.0264, -0.0127$), captures a longer-term structural trend that is not visually apparent from the raw data. This cycle likely reflects the gradual phase-out of older reactors, extended plant life extensions, and the commissioning of new units over the past two decades. The negative sign of $\beta_5$ suggests a declining secular trend in nuclear generation over this 9-year horizon, consistent with the broader context of the U.S. nuclear industry, which has faced economic challenges from cheaper natural gas and renewables, leading to premature retirements of several facilities. The estimated dispersion parameter $\hat\nu = 770.75$ indicates that the Gamma-GLARMA model adequately captures the mean structure, with relatively low overdispersion relative to the scale of generation (which is in the order of thousands of GWh). From a forecasting perspective, the significant seasonal components provide a reliable baseline for predicting nuclear generation up to 12 months ahead, whereas the long-term cycle offers guidance for infrastructure planning and policy assessment over the next decade. The presence of a negative AR(12) coefficient ($\hat\phi_{12} = -0.0124$), although small in magnitude, suggests a slight dampening effect from the same month of the previous year, which may be attributed to year-over-year adjustments in maintenance scheduling. Overall, the model not only fits the historical data well but also uncovers meaningful temporal patterns that can inform both short-term grid management and long-term energy policy decisions regarding the role of nuclear power in the U.S. energy mix.

\subsection{Air Pollution and hospital admissions by COPD}

This application explores the association between air pollutants and the monthly number of Chronic Obstructive Pulmonary Disease (COPD) cases in Belo Horizonte, Brazil, from 2007 to 2013 ($n=84$). The primary objective is to model the response variable while addressing the inherent challenges of count data, such as overdispersion, seasonality, and multicollinearity among predictors.

\begin{figure}[!ht]
    \centering
    \includegraphics[width=0.5\textwidth]{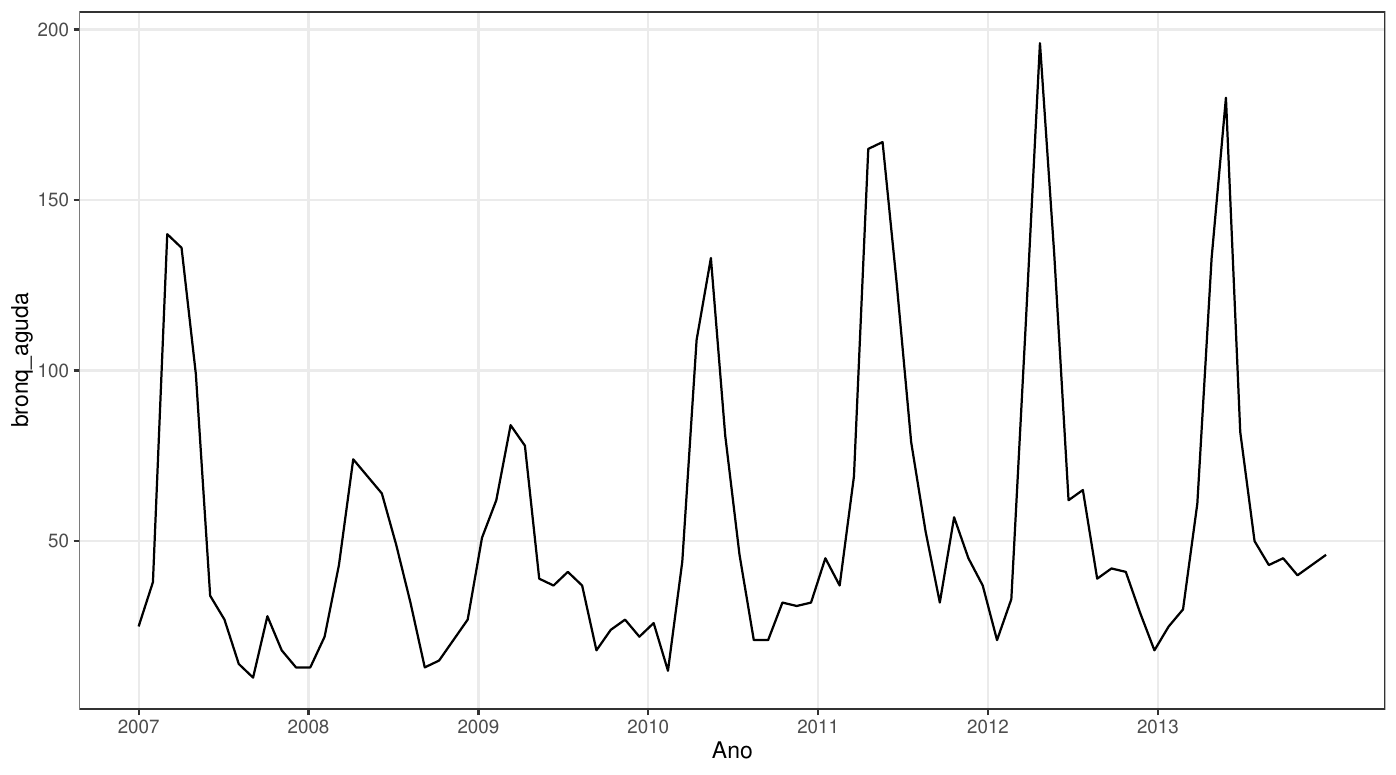}
    \caption{Time series of monthly COPD cases}
    \label{fig:copd_cases}
\end{figure}

\subsubsection{Dimensionality Reduction: Principal Component Analysis}

Initial exploratory analysis revealed significant correlation among air pollutants ($PM_{10}$, $NO$, $NO_2$, $CO$, $NO_X$, and $O_3$), suggesting that a direct regression approach would suffer from multicollinearity. To overcome this, Principal Component Analysis (PCA) was applied to the standardized pollutants.
\\
\begin{figure}[!ht]
    \centering
    \includegraphics[width=\textwidth]{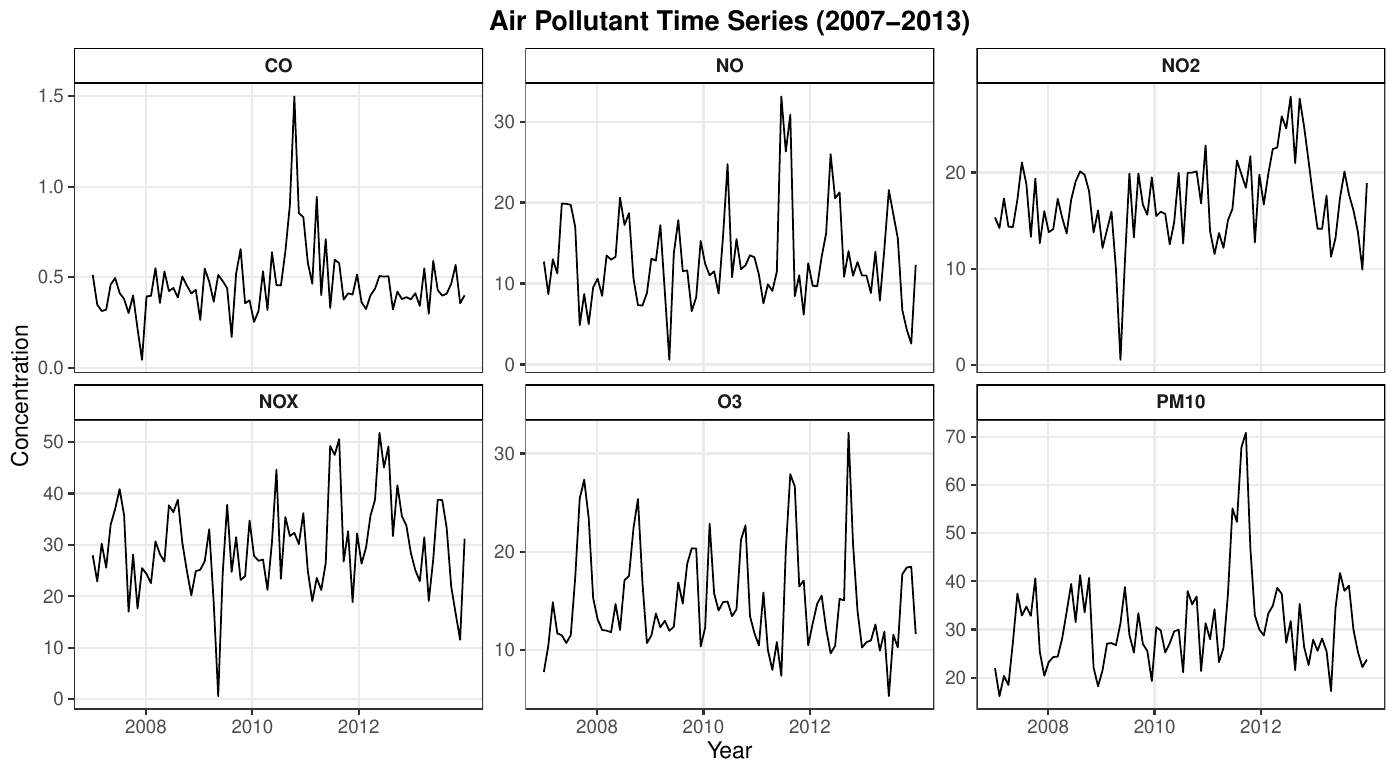} 
    \caption{Monthly concentrations of air pollutants in Belo Horizonte from January 2007 to December 2013.}
    \label{fig:pollutants}
\end{figure}

Table \ref{tab:pca_variance} summarizes the importance of the components. The first three principal components (PC1, PC2, and PC3) explain approximately $84.57\%$ of the total variance. Specifically, PC1 alone accounts for $47.42\%$, being strongly influenced by nitrogen oxides ($NO$ and $NO_X$), which are typical markers of vehicular combustion.

\begin{table}[!ht]
\centering
\caption{Variance explained by the Principal Components of air pollutants.}
\label{tab:pca_variance}
\begin{tabular}{lrrr}
\hline
Component & Proportion of Variance (\%) & Cumulative Proportion (\%) \\ \hline
PC1       & 47.42                      & 47.42                     \\
PC2       & 21.43                      & 68.85                     \\
PC3       & 15.72                      & 84.57                     \\
PC4       & 11.89                      & 96.47                     \\
PC5       &  3.53                      & 99.99                     \\      PC6       &  0.01                      & 100.00                     \\ 
\hline
\end{tabular}
\end{table}

\subsubsection{Model Specification and Bayesian Inference}

We implemented a Bayesian GLARMA model with a Negative Binomial distribution to account for the observed overdispersion in the COPD cases. The linear predictor is defined as follows:
\begin{equation*}
\log(\mu_t) = \alpha + \sum_{j=1}^{3} \beta_j PC_{j,t} + b_1 \sin\left(\frac{2\pi t}{12}\right) + b_2\cos\left(\frac{2\pi t}{12}\right) + b_3 \sin\left(\frac{2\pi t}{6}\right) + b_4 \cos\left(\frac{2\pi t}{6}\right) + Z_t,
\end{equation*}
where $Z_t = \phi (Z_{t-1} + e_{t-1})$ represents the autoregressive structure driven by the dynamic Pearson residuals. 

The Bayesian estimation via MCMC is conducted using weakly informative prior distributions to ensure numerical stability. Specifically, we assign a normal prior centered near the log-mean of the response variable for the intercept, $\alpha \sim N(4, 2^2)$, a continuous uniform prior over the stationary domain for the autoregressive parameter, $\phi \sim U(-1, 1)$, and a flexible exponential distribution for the negative binomial dispersion parameter, $r \sim U(0,50)$. For the structural regression coefficients associated with the three principal components, $\beta_j$ (for $j=1,2,3$), as well as for the four harmonic amplitudes, $b_k$ (for $k=1,2,3,4$), independent standard normal prior distributions are specified, $\beta_j, b_k \sim N(0, 1)$. 

Posterior inference is performed using the \texttt{nimble} package by running $2$ independent Markov chains of $10,000$ iterations each, applying a $30\%$ burn-in period to discard transient initial states. Chain convergence and mixing properties were formally confirmed through the Gelman-Rubin diagnostic ($\hat{R} \approx 1.01$ across all parameters) alongside visual inspection of trace plots and autocorrelation functions.

\subsubsection{Results}

The posterior summaries are presented in Table \ref{tab:posterior_results}. A significant positive effect was found for PC1 ($\hat{\beta}_1 = 0.075$, $95\%$ HPD: $[0.033, 0.118]$), with a posterior probability of $P(\beta_1 > 0) = 99.89\%$. This suggests that increases in the concentration of pollutants associated with PC1 are strongly related to a rise in COPD cases. PC2 also showed a high probability of a positive effect ($89.3\%$).

\begin{table}[!ht]
\centering
\caption{Posterior mean, standard deviation (SD), and 95\% Highest Posterior Density (HPD) intervals for the GLARMA model parameters.}\vspace{.3cm}
\label{tab:posterior_results}
\begin{tabular}{lrrcc}
\hline
Parameter & Mean & SD & 95\% HPD Lower & 95\% HPD Upper \\ \hline
$\alpha$  & 3.826 & 0.073 & 3.694 & 3.968 \\
$\beta_1$ & 0.075 & 0.022 & 0.033 & 0.118 \\
$\beta_2$ & 0.050 & 0.041 & -0.026 & 0.129 \\
$\beta_3$ & 0.011 & 0.034 & -0.052 & 0.074 \\
$\phi_1$    & 0.251 & 0.045 & 0.173 & 0.345 \\
$r$ & 12.223 & 2.462 & 7.796 & 17.102 \\
$b_1$     & 0.449 & 0.102 & 0.259 & 0.669 \\
$b_2$     & -0.435 & 0.095 & -0.622 & -0.235 \\
$b_3$     & -0.346 & 0.073 & -0.483 & -0.199 \\
$b_4$     & -0.066 & 0.080 & -0.217 & -0.088 \\
\hline
\end{tabular}
\end{table}

The harmonic coefficients indicated significant annual and semiannual seasonality, as their HPD intervals do not include zero. The autoregressive parameter $\phi_1 \approx 0.25$ confirms the existence of serial dependence in the data, which is successfully captured by the GLARMA framework.

\subsubsection{Residual Diagnostics}

\begin{figure}[!ht]
\centering
\includegraphics[width=0.9\textwidth]{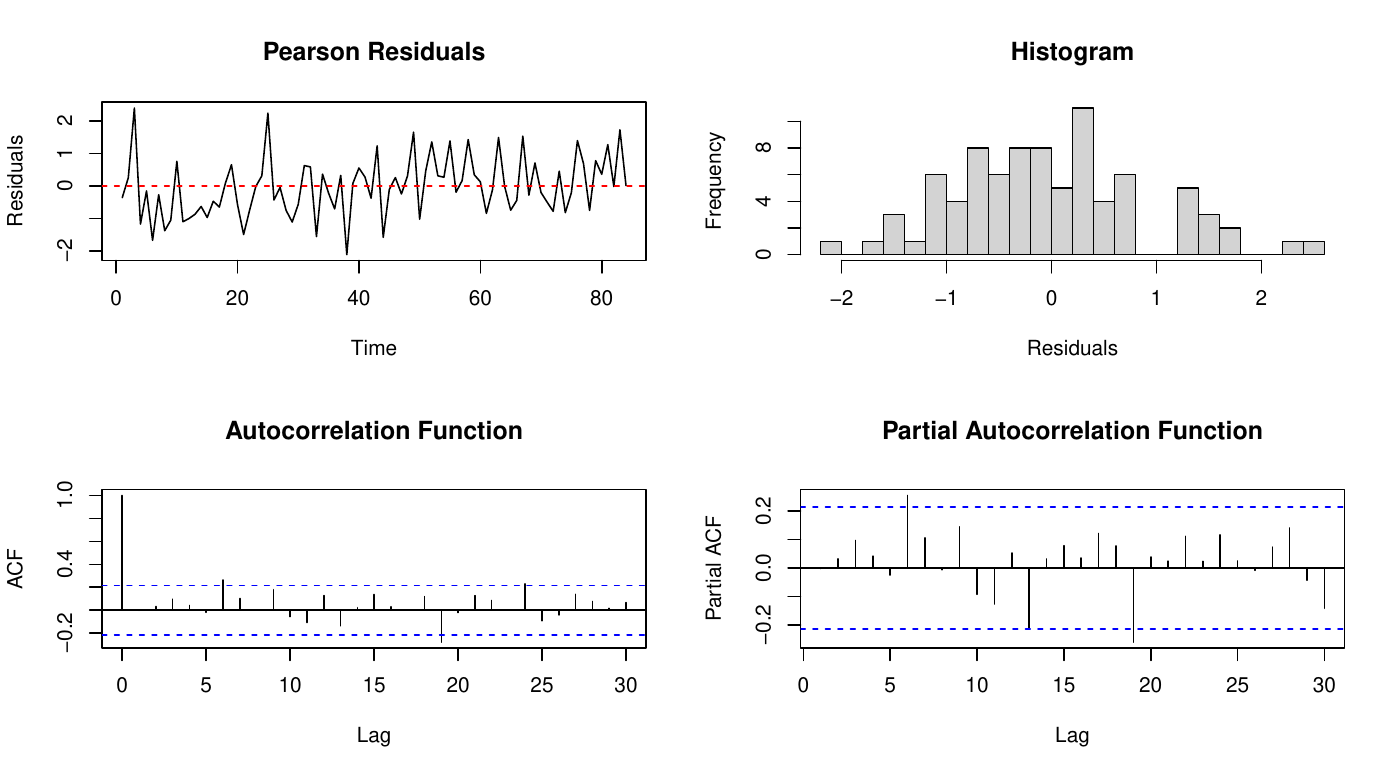}
\caption{Diagnostic plots for Pearson residuals.}
\label{fig:residuals}
\end{figure}

The model's goodness-of-fit was assessed through a comprehensive analysis of the Pearson residuals, as illustrated in Figure \ref{fig:residuals}. The time plot (top-left panel) shows that the residuals fluctuate randomly within a stable horizontal band centered at zero, indicating a satisfactory structural fit and homoscedasticity. The histogram (top-right panel) exhibits an approximatelly symmetric, bell-shaped distribution centered around zero, supporting the assumption of asymptotic normality. A minor discontinuity or ``gap'' is visible near the center of the distribution. This is a typical finite-sample phenomenon in count time series ($n=84$), arising because the original discrete observations ($Y_t$) are combined with continuous conditional means ($\mu_t$), which naturally restricts the residuals from taking certain intermediate values. 

The absence of autocorrelation was confirmed by the Ljung-Box test ($Q = 11.95, p = 0.28$ for lag 10), indicating that the residuals behave as white noise. Furthermore, the Phillips-Perron test ($p = 0.01$) rejected the null hypothesis of a unit root, supporting the stationarity of the residuals. In addition, the Autocorrelation Function (ACF) plot in Figure \ref{fig:residuals}, (bottom-left panel), presents lags beyond zero fall entirely within the 95\% confidence bands. This lack of serial dependence is supported by the Partial Autocorrelation Function (PACF) plot (bottom-right panel).

\begin{figure}[!ht]
\centering
\includegraphics[width=0.6\textwidth]{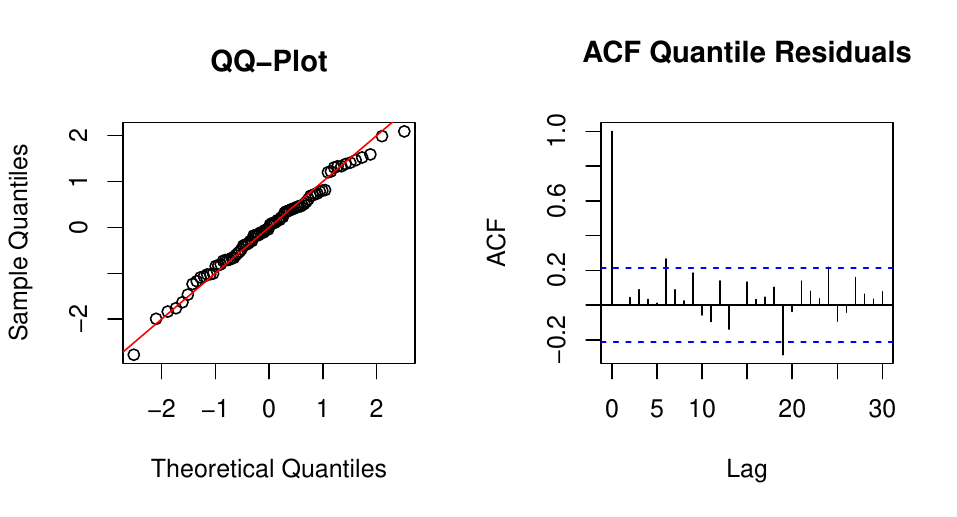}
\caption{Diagnostic plots for Quantile residuals.}
\label{fig:quantile_res}
\end{figure}

To overcome the inherent limitations of Pearson residuals in discrete settings, we implemented randomized quantile residuals. Once the original data consists of counts, these residuals map the discrete probabilities into a continuous scale that should asymptotically follow a standard normal distribution under correct model specification. Figure \ref{fig:quantile_res} presents the diagnostic plots for these quantile residuals. The QQ-Plot displays a remarkable alignment of the sample quantiles along the theoretical $45^\circ$ reference line, showing no heavy tails or severe outliers. This visual behavior is formally backed by the Shapiro-Wilk normality test, which yielded a statistic of $W = 0.991$ and a highly non-significant $p$-value of $0.841$, strongly supporting the assumption of normality and confirming that the Negative Binomial distribution perfectly accommodates the overdispersion of the monthly COPD cases. Furthermore, the ACF plot demonstrates that the temporal dependencies are virtually fully resolved. Although a single isolated lag marginally crosses the boundary, this minor fluctuation is statistically expected under a 5\% significance level due to multiple testing, without indicating model inadequacy. Combined with the highly non-significant Ljung-Box test ($Q = 11.95$, $p = 0.28$), these results structurally support that the proposed Bayesian framework effectively reduced the errors to a pure white noise process.
%

\section{Conclusion}

In this paper, we introduced a comprehensive Bayesian framework for the estimation and inference of general GLARMA models, extending the methodology beyond the common exponential family setting. Our approach successfully handles positive, double-bounded, and count time series through a unified MCMC-based estimation procedure. The extensive Monte Carlo simulation study demonstrated that the proposed methodology yields reliable and stable parameter estimates across a range of distributional assumptions and persistence levels, including challenging scenarios near the boundary of the stationarity region. The empirical applications to the U.S. nuclear energy generation data and COPD hospital admissions illustrated the practical utility and flexibility of the framework in real-world settings, uncovering meaningful covariate effects and temporal patterns. A key limitation of our study is the heuristic nature of the model selection procedure, which relies on posterior diagnostics and residual analysis. Future work in this direction could explore other methods for model selection, such as Reversible Jump Markov Chain Monte Carlo and spike-and-slab priors. We believe that the proposed framework represents a significant step towards making Bayesian inference for general GLARMA models a practical and accessible tool for applied researchers.

\subsection*{Acknowledgments}
G. Pumi  gratefully acknowledges the financial support received by the Conselho Nacional de Desenvolvimento Cient\'ifico e Tecnol\'ogico -- CNPq Brasil  -- Bolsa de Produtividade em Pesquisa - Proc. 303281/2025-1. \\A.J. Camara gratefully acknowledges the financial support received by the Fundação de Amparo à Pesquisa e Inovação do Espírito Santo -- FAPES -- Edital FAPES Universal 13/2025 - Proc. 2025-KZNQK.

\subsection*{Conflict of Interest Statement}
The authors declare no conflicts of interest.

\bibliographystyle{apalike} 
\bibliography{ref}

\end{document}